\documentclass[epj]{webofc}

\usepackage[varg]{txfonts}   
\usepackage{subfigure}
\usepackage{hyperref}

\newcommand{\br}[1]{\langle #1\rangle}

\woctitle{XLVI International Symposium on Multiparticle Dynamics}
\begin{document}

\title{Longitudinal correlations in the initial stages of ultra-relativistic nuclear collisions}

\author{
        Wojciech Broniowski\inst{2,3}\fnsep\thanks{\email{Wojciech.Broniowski@ifj.edu.pl}}\and Piotr Bo\.{z}ek\inst{1}\fnsep\thanks{\email{Piotr.Bozek@fis.agh.edu.pl}} 
}

\institute{AGH University of Science and Technology, Faculty of Physics and
Applied Computer Science, \\ 30-059~Cracow, Poland 
\and
The H. Niewodnicza\'{n}ski Institute of Nuclear Physics,
Polish Academy of Sciences, 31-342 Cracow, Poland
\and
Institute of Physics, Jan Kochanowski University, 25-406 Kielce, Poland
          }

\abstract{In this talk we review some of our results for the longitudinal correlations in connection with recent experimental 
data, in particular for the {\em torque} effect and for the two-particle correlations in pseudorapidity, $C(\eta_1,\eta_2)$.
The model framework involves event-by-event fluctuations of the initial conditions, followed with 3D viscous hydrodynamic
evolution and statistical hadronization.}

\maketitle

The phenomenon of collectivity allows for simple understanding of very basic features found in 
ultra-relativistic nuclear collisions (for a review and literature see, e.g., \cite{Heinz:2013th}), 
such as the harmonic flow, the emergence of the near-side ridge, or even 
the transverse-momentum fluctuations~\cite{Broniowski:2009fm,Bozek:2012fw}. 

\begin{figure}[tb]
\begin{center}
\includegraphics[angle=90,width=.67 \textwidth]{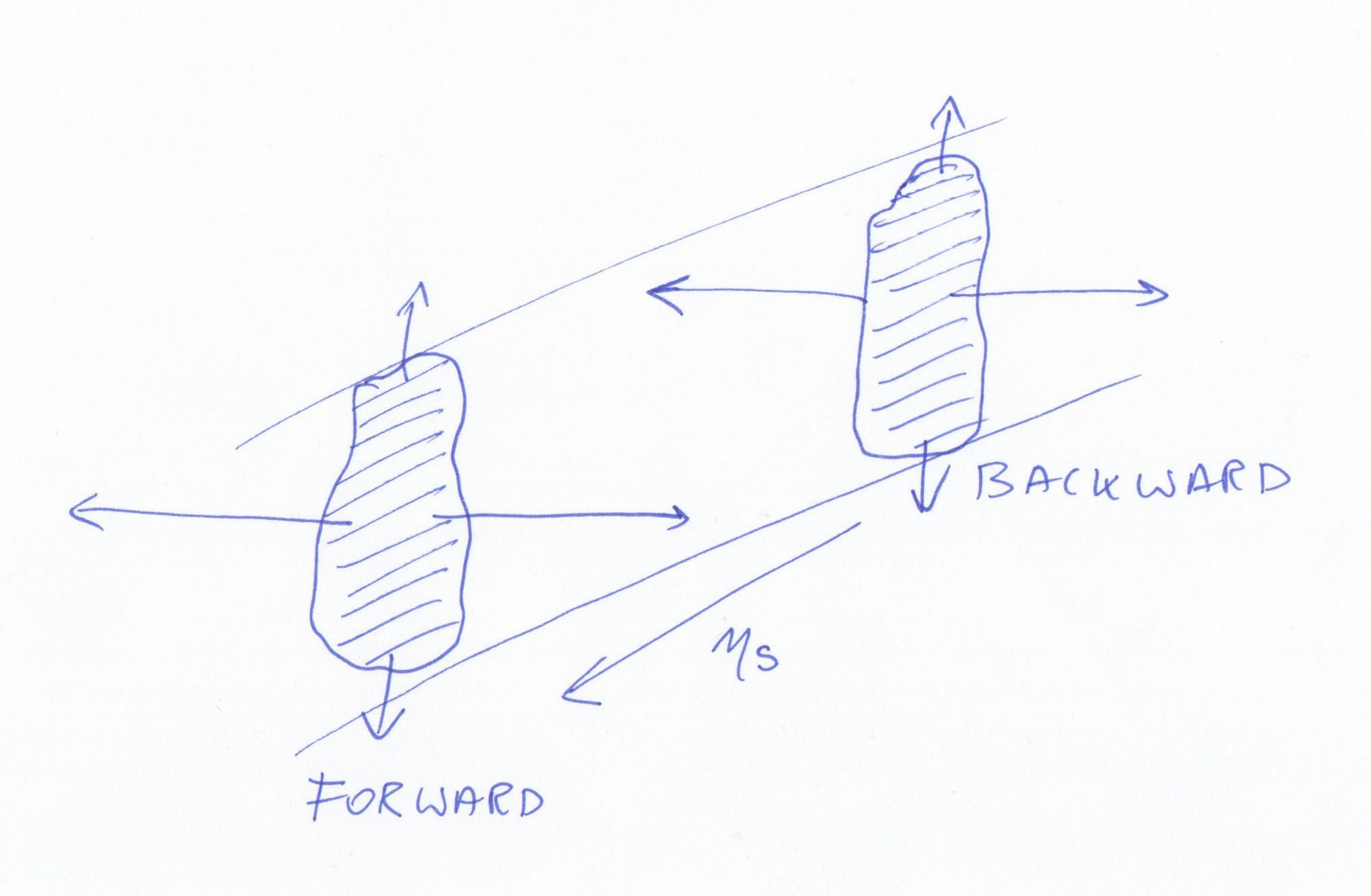} 
\caption{Collimation of flow at very distant longitudinal separations, following from the shape-flow transmutation. \label{fig:collim}}
\end{center}
\end{figure}

In the long-range rapidity correlations, the phenomenon is manifest through the shape-flow transmutation
depicted in Fig.~\ref{fig:collim}: If the transverse sections in the initial state are similar in the forward (F) and backward (B) directions, then the 
corresponding harmonic flow (directions of the principal axes, magnitude) are also similar, up to decorrelation effects from 
fluctuations. Of course, the similarity of the F and B shapes must come out from the early production mechanism, for instance flux-tubes
spanned along a large rapidity range. Thus, actually, the starting point of the approach relies on an exact similarity of the transverse fireball 
shapes in distant F and B directions, which would lead to perfect collimation of the flow axes.

\begin{figure}[tb]
\begin{center}
\includegraphics[angle=0,width=.3 \textwidth]{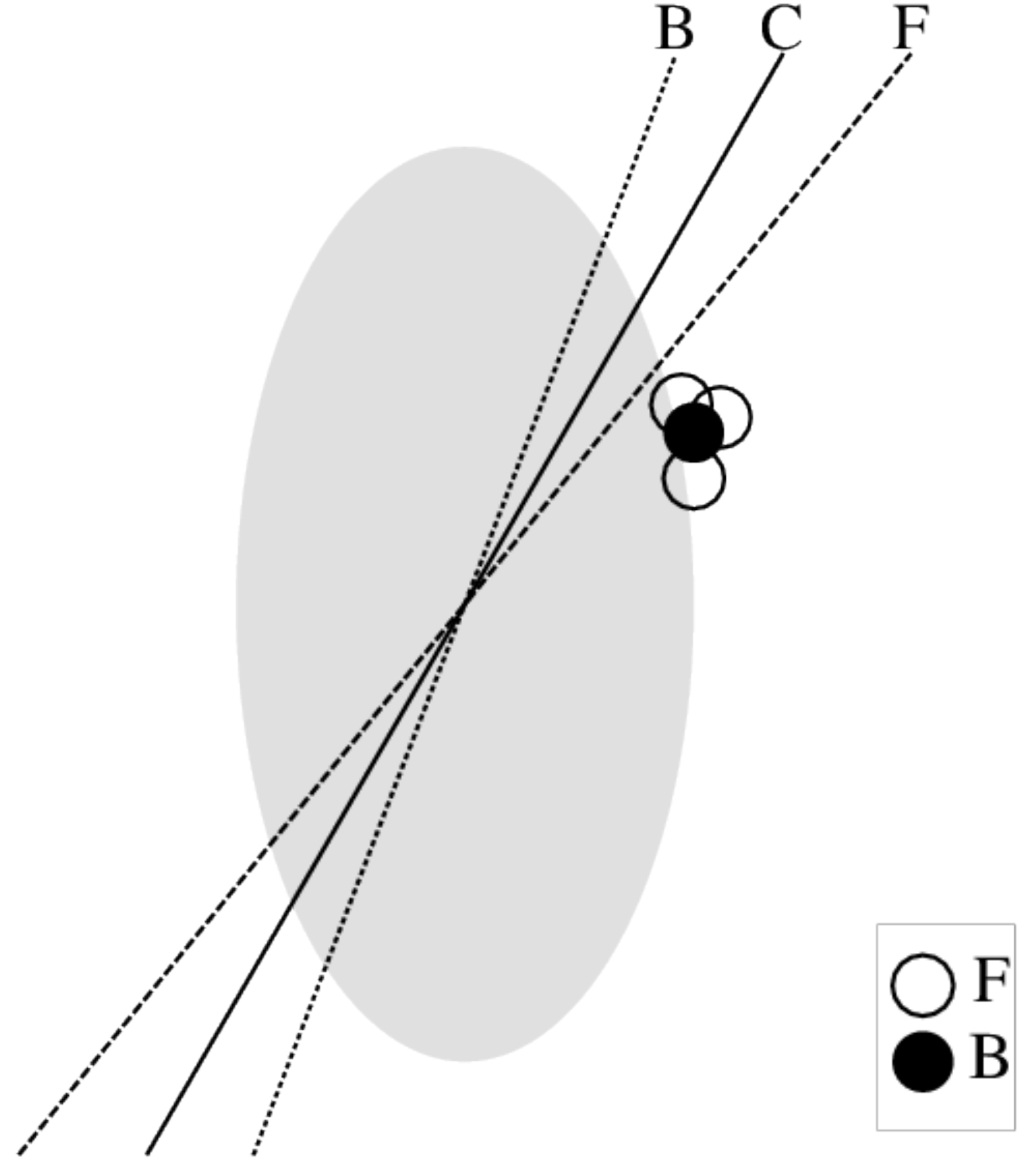} \qquad \includegraphics[angle=0,width=.44 \textwidth]{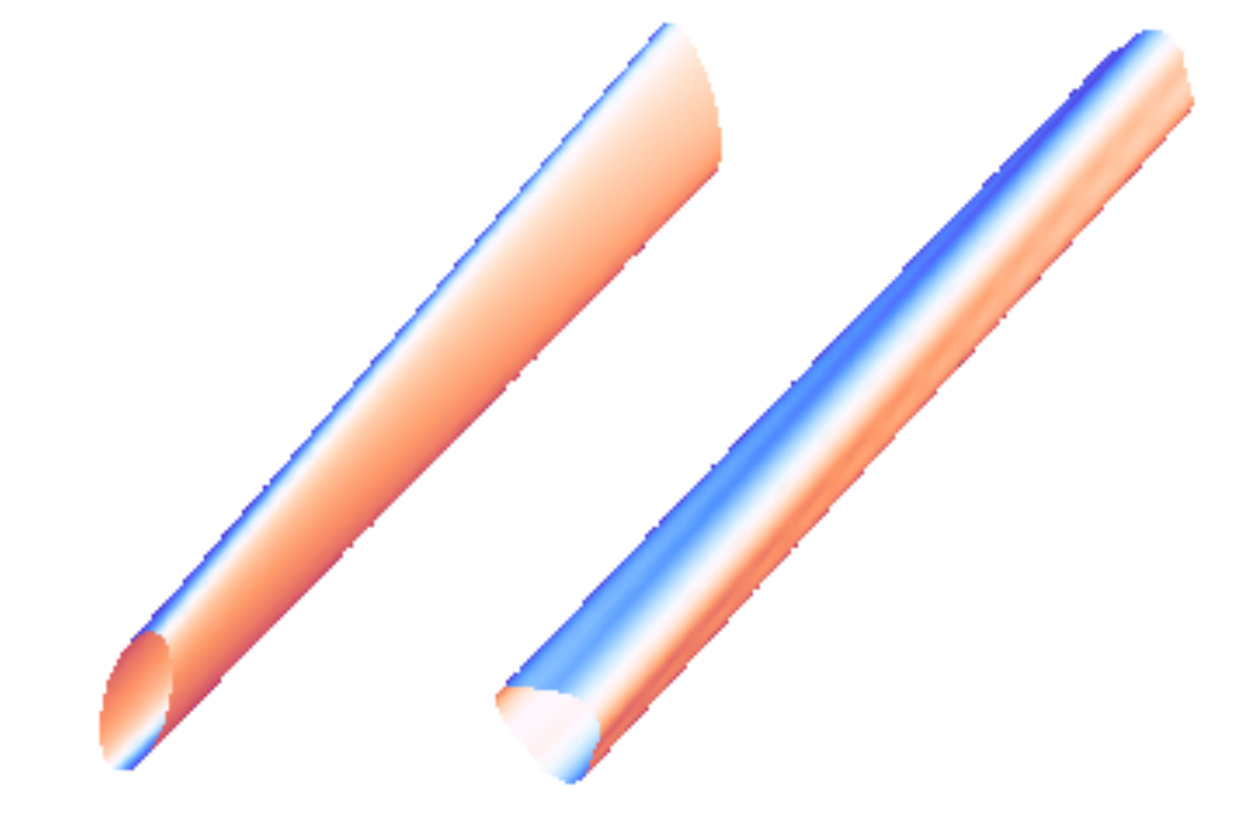}
\caption{The torque effect. The fluctuations (left) in the number of forward- (F) and backward-going (B) participants, together 
with asymmetric emission profiles in rapidity, lead to event-by-event fluctuations of the event-plane angle (right).~\cite{Bozek:2010vz} \label{fig:torque}}
\end{center}
\end{figure}

The torque effect, cartooned in Fig.~\ref{fig:torque}, quantifies departure from this perfect collimation. As originally proposed 
in Ref.~\cite{Bozek:2010vz} in the framework of the wounded-nucleon model~\cite{Bialas:1976ed}, the decollimation happens 
as a combination of two features:
\begin{itemize}
\item First, there are event-by-event fluctuations in the number of the F and B going participants (left part of Fig.~\ref{fig:torque}). 
\item Second, the participants must have 
asymmetric emission profiles (relating to the deposition of the initial entropy) in the rapidity, with a preference that a 
participant shines preferably in its forward direction~\cite{Bialas:2004su}. 
\end{itemize}
These lead to the event-by-event torque, as 
shown in the right part of  Fig.~\ref{fig:torque}, appearing independently for subsequent Fourier components.

Apart for the decollimation due to the initial fluctuations, which are the object of our interest, there are also ``unwanted'' final-stage 
fluctuation resulting from the statistical nature of hadronization. These may be eliminated through the use of appropriate 
measures, for instance the 3-bin correlator proposed by the CMS Collaboration~\cite{Khachatryan:2015oea}:
\begin{eqnarray}
&& r_n(\eta^a, \eta^b)=\frac{V_{n\Delta}(-\eta^a,\eta^b)}{V_{n\Delta}(\eta^a,\eta^b)}.  \\
&&  V_{n\Delta}(\eta^a,\eta^b) = \langle \langle e^{i n(\phi_1-\phi_2)} \rangle \rangle 
= \langle \langle e^{i n(\psi_n(\eta_a)+\phi'_1-\psi_n(\eta_b)-\phi'_2)} \rangle \rangle 
\simeq  \langle e^{i n[\psi_n(\eta_a)-\psi_n(\eta_b)]} \rangle \langle \langle e^{i n \phi'_1 -i n \phi'_2} \rangle \rangle. \nonumber \label{eq:fact}
\end{eqnarray}
The azimuthal angles of the measured particles, $\phi_1$ and $\phi_2$, are obtained in a certain
reference frame, whereas $\psi_n(\eta_a)$ and $\psi_n(\eta_b)$ are the event-plane angles of the fireball in bins
centered around $\eta_a$ and $\eta_b$. The angles 
$\phi'_1$ and $\phi'_2$ are evaluated relative to $\psi_{n}(\eta_a)$ and $\psi_{n}(\eta_b)$, respectively. 
The forward reference bin $4,4< \eta_b < 5$ is well  separated from the two measurement 
bins centered at $\eta_a$ and $-\eta_a$. 

\begin{figure}[tb]
\begin{center}
\includegraphics[angle=0,width=.45 \textwidth]{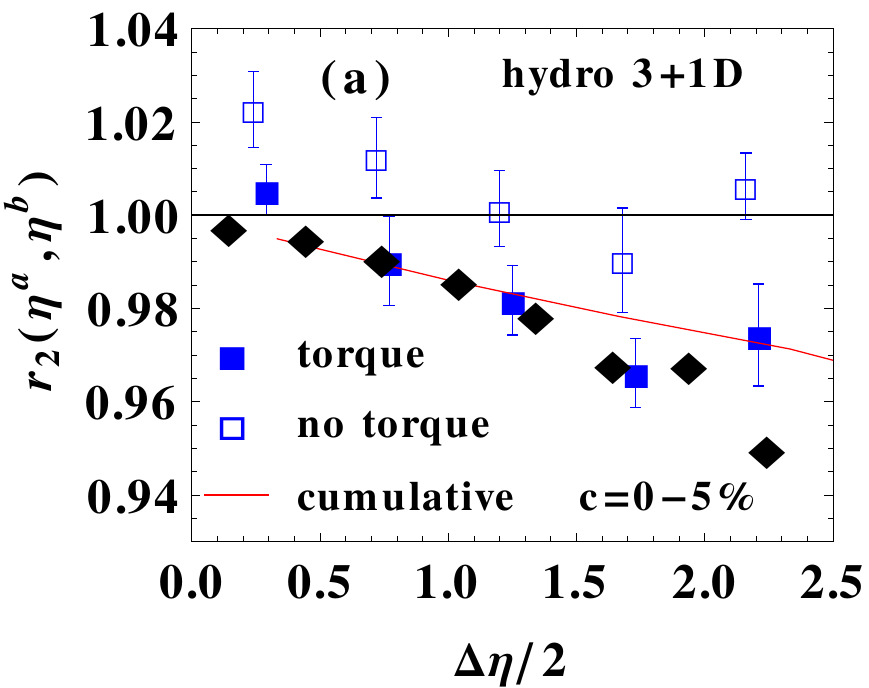} \hfill \includegraphics[angle=0,width=.45 \textwidth]{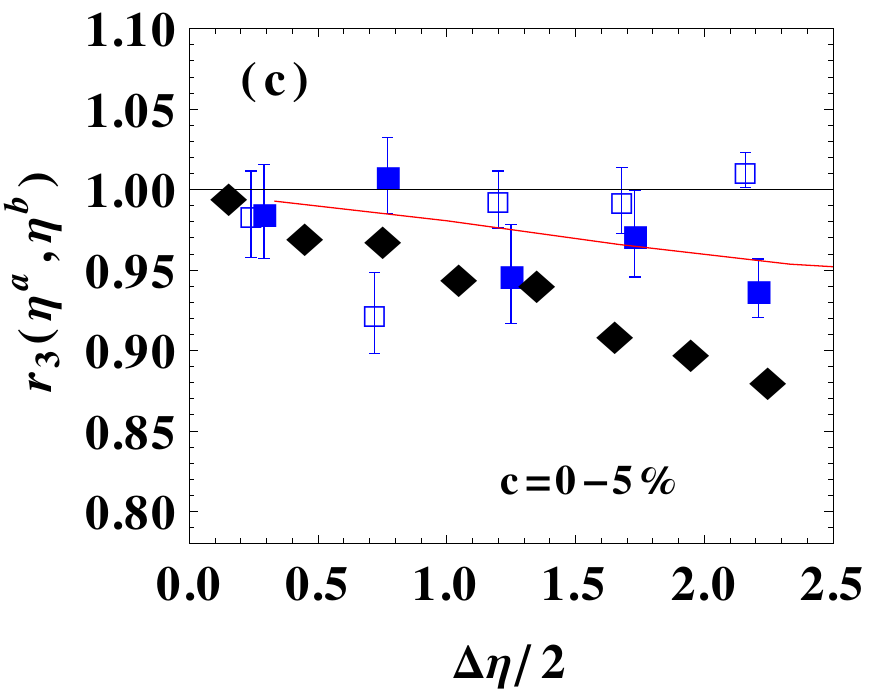}
\caption{The torque effect for Pb+Pb collisions at $\sqrt{s_{NN}}=2.76$~TeV, 
expressed through the 3-bin CMS measure $r_n(\eta^a,\eta^b)$, plotted as a function of $\Delta \eta/2 \equiv \eta_a$ 
with the reference bin $4<\eta_b<5$. 
Our event-by-event model simulations~\cite{Bozek:2015bha} qualitatively reproduce the CMS data~\cite{Khachatryan:2015oea} when the ingredients leading to
torque are incorporated (in particular, the asymmetric particle emission profile from a given wounded nucleon). 
The effect is significantly larger for the triangular flow (right) than for the elliptic flow (left). \label{fig:LHC}}
\end{center}
\end{figure}

The departure of $r_n$ from unity, seen in Fig.~\ref{fig:LHC}, is a measure of 
the event-plane angle decorrelation, i.e., the torque. We note that our simulations, 
based on GLISSANDO~\cite{Broniowski:2007nz,Rybczynski:2013yba}, 3D viscous 
hydrodynamics~\cite{Bozek:2011ua}, and THERMINATOR~\cite{Kisiel:2005hn,Chojnacki:2011hb}, lead to fair agreement with the data
when the two conditions itemized above are incorporated (filled 
points labeled ``torque'' in Fig.~\ref{fig:LHC}). Without asymmetric production profiles there is no effect (empty squares labeled ``no torque'').
The solid lines in Fig.~\ref{fig:LHC} show the calculation where many THERMINATOR events for the same initial condition are averaged (cumulative events) 
to improve statistics.

\begin{figure}[tb]
\begin{center}
\includegraphics[angle=90,width=.39 \textwidth]{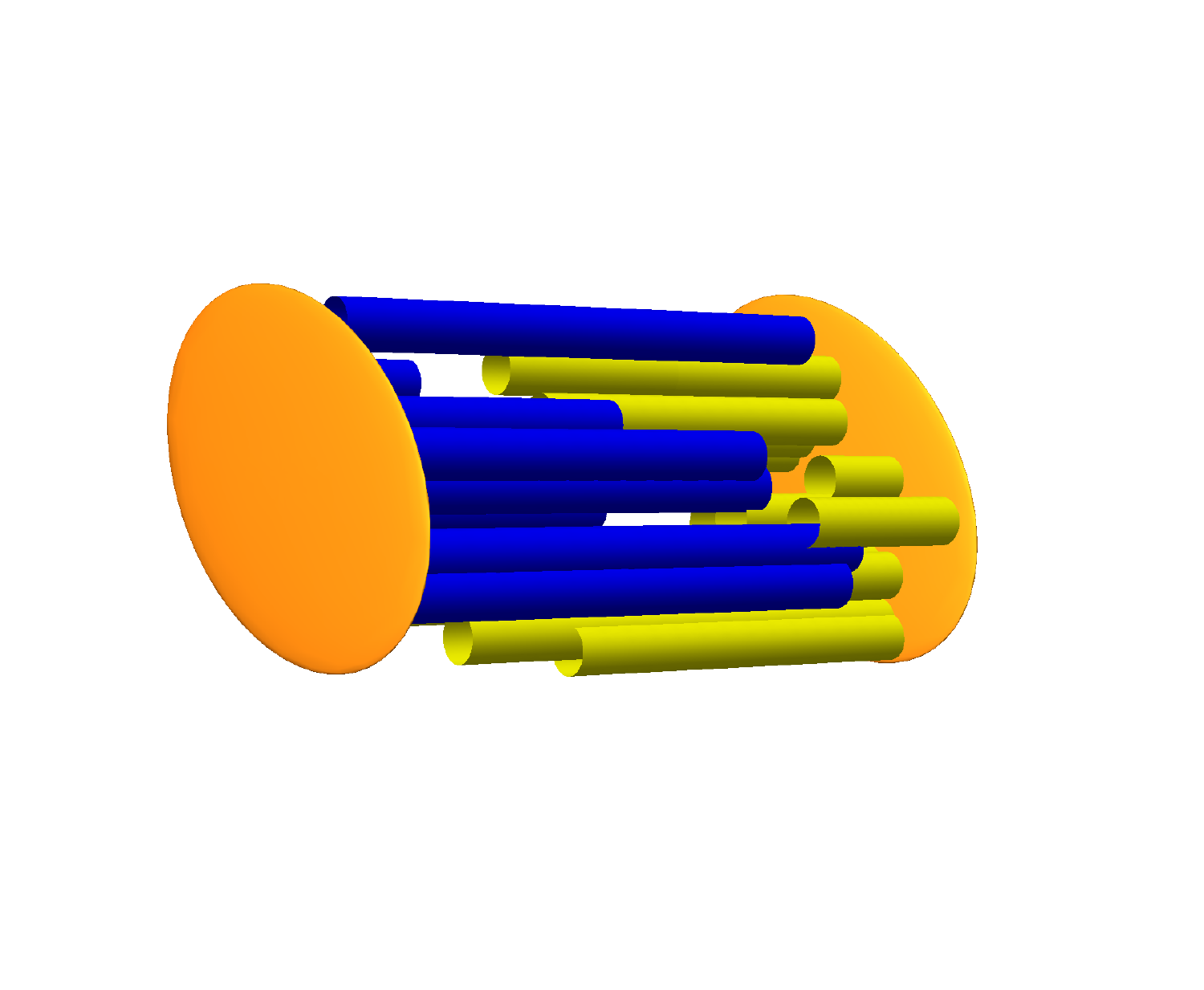} \hfill \includegraphics[angle=0,width=.6 \textwidth]{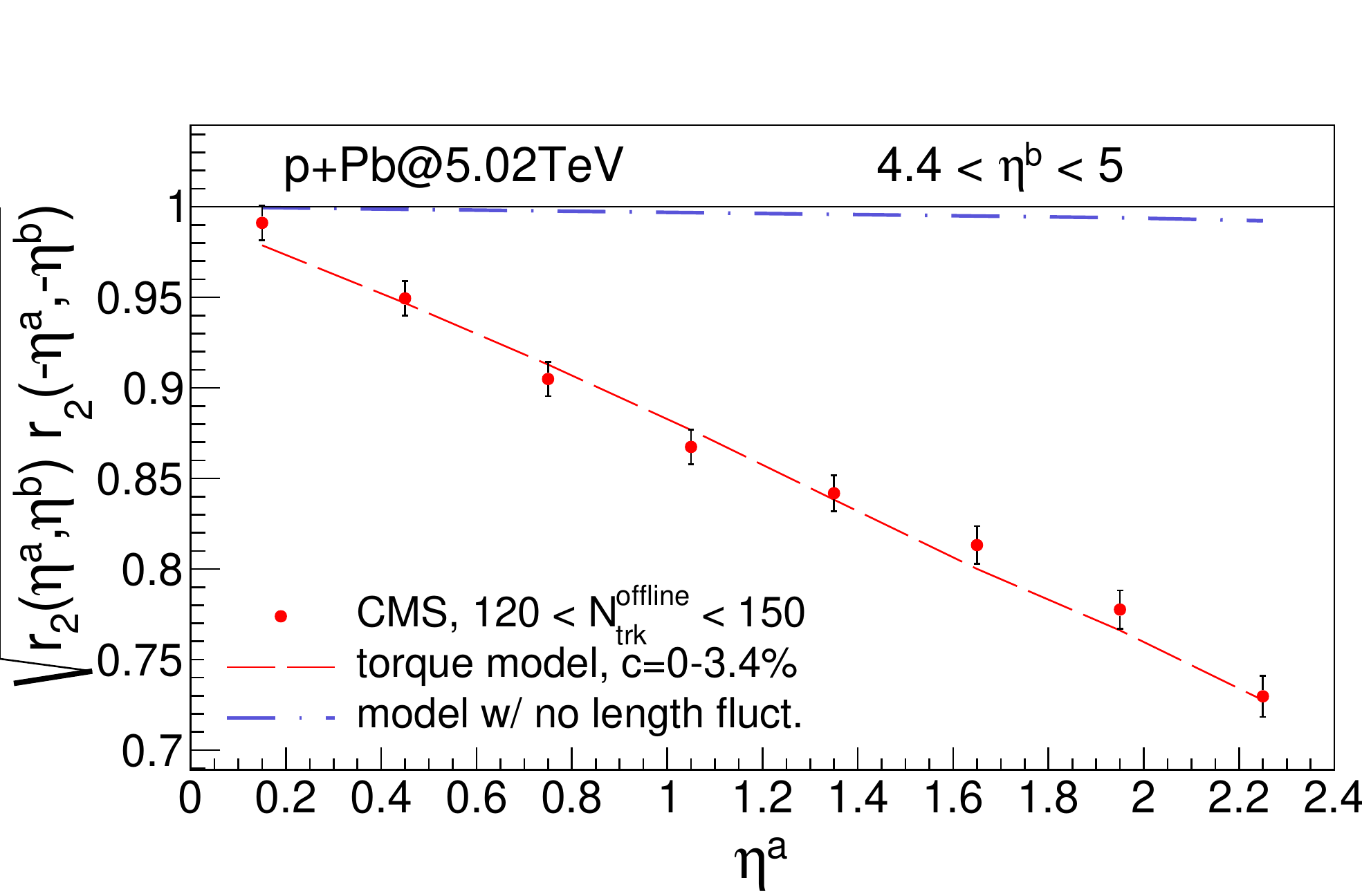}
\caption{The torque effect for p-Pb. \label{fig:pPb} The decorrelation mechanism from string breaking (left) is crucial for the 
explanation of the CMS data~\cite{Khachatryan:2015oea}. For strings spanning all the pseudorapidity range there is no 
decorrelation (dash-dotted line in the right part). The model calculation is carried out with GLISSANDO~\cite{Rybczynski:2013yba}.}
\end{center}
\end{figure}

Interestingly, the CMS measurements for the p-Pb collisions~\cite{Khachatryan:2015oea} indicates the necessity of an additional decorrelation 
mechanism:
\begin{itemize}
 \item Fluctuating strings.
\end{itemize}
This is because in p-A collisions the transverse shape of the fireball is determined by the positions of the participants in A. 
If the corresponding strings span all the relevant range in pseudorapidity, the transverse shape of the fireball is the same at all pseudorapidities 
(see Ref.~\cite{Bozek:2015bna} for a detailed discussion). Adding fluctuating end-points of the strings~\cite{Brodsky:1977de,Bialas:2004kt} 
(see the left part of Fig.~\ref{fig:pPb}) leads to proper description of the data.

\begin{figure}[tb]
\begin{center}
\includegraphics[angle=0,width=.45 \textwidth]{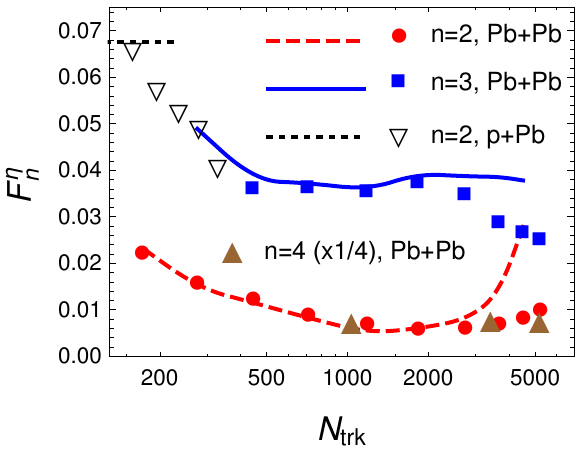}
\caption{Slope in $\eta_a$ of the $r_n$ decorrelation measure for Pb-Pb and p-Pb collisions. The data are from 
Ref.~\cite{Khachatryan:2015oea}, and the model results are from GLISSANDO~\cite{Rybczynski:2013yba}. 
We note fair agreement, except for the highest multiplicity events. Note that relation~(\ref{eq:F}) is well satisfied by the data. \label{fig:F2}}
\end{center}
\end{figure}

In Fig.~\ref{fig:F2} we summarize the results for the slope parameter introduced by the CMS collaboration,
\begin{equation}
r_n(\eta_a,\eta_b) = e^{-2 F^\eta_n \eta_a}.
\label{eq:F}
\end{equation}
We note that from general considerations~\cite{Bozek:2015bna} the slope for the $n=4$ harmonic is very simply related to 
the slope for $n=2$:
\begin{eqnarray}
F^\eta_4/4\simeq F^\eta_2.
\label{eq:F4}
\end{eqnarray}

\begin{figure}[tb]
\begin{center}
\includegraphics[angle=0,width=.42 \textwidth]{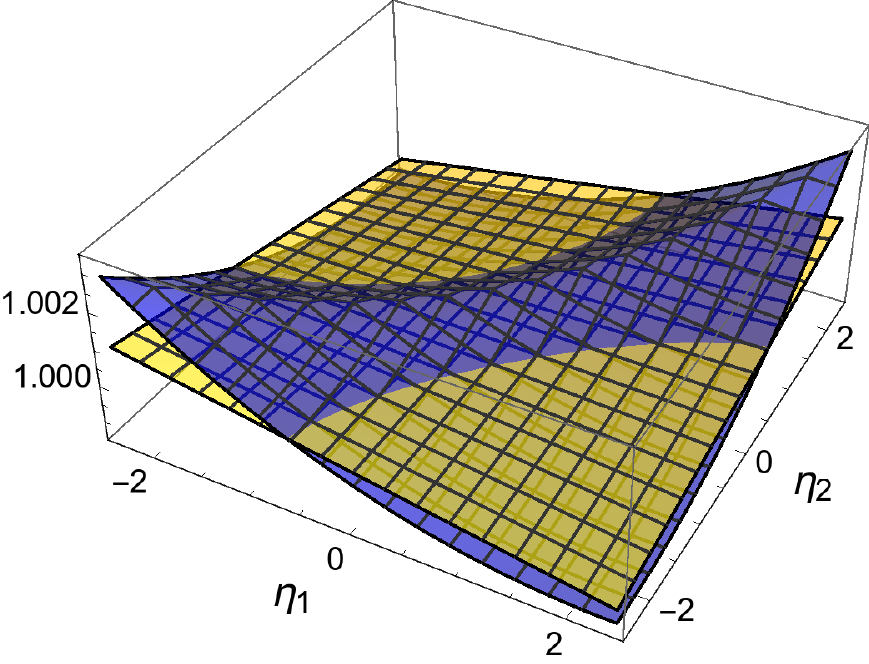} \includegraphics[angle=0,width=.5 \textwidth]{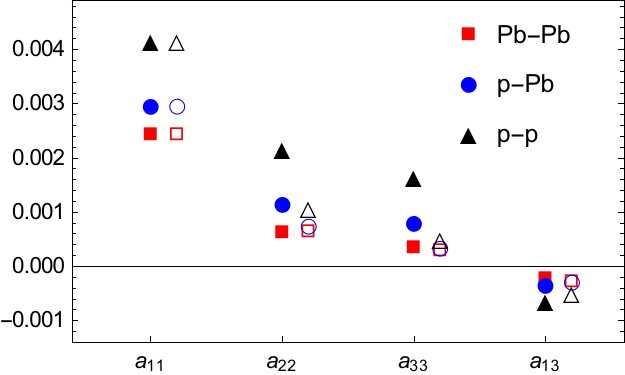}
\caption{Left: Model prediction for the normalized correlation function ${C}_N(\eta_1,\eta_2)$ for Pb-Pb collisions at the LHC for $c=30-40\%$. 
The lighter color surface corresponds to the calculation without string fluctuations. The surface with an elongated ridge (darker color) 
corresponds to the case incorporating the string  fluctuations. Right: the $a_{nm}$ coefficients  \label{fig:anm}}
\end{center}
\end{figure}

Next, we discuss the interpretation of the recent ATLAS measurements~\cite{ATLAS:2015kla,ATLAS:anm} of the two-particle correlation 
function in pseudorapidity, $C_N(\eta_1, \eta_2)$ (for details see Ref.~\cite{Broniowski:2015oif}). A typical model result is shown in the 
left part of Fig.~\ref{fig:anm}, where we see that the inclusion of the string fluctuations enhances the correlations and builds-up a characteristic
ridge structure along $\eta_1=\eta_2$. Quantitative measures are conveniently provided with the $a_{nm}$ coefficients defined in Refs,~\cite{Bzdak:2012tp,Jia:2015jga}.
In our simple model the corresponding formulas are analytic~\cite{Broniowski:2015oif}. We introduce  $N_+=N_A+N_B$, $N_-=N_A-N_B$, 
where $N_{A,B}$ denotes the number of sources associated with the given nucleus.
We obtain the non-vanishing terms (for $n,m>0$)
\begin{eqnarray}
a_{nn}&=& \frac{{{\rm var}(N_-)}/{\br{N_+}}+(1-r)s(\omega)-r}{6 \br{N_+}}\frac{Y^2}{y_b^2} \delta_{n1} 
+ r \frac{s(\omega)+1}{(2n-1)(2n+3) \br{N_+}}\frac{Y}{y_b}, \label{eq:ann} \\  \nonumber \\
a_{n,n+2} &=&a_{n+2,n} = -r \frac{s(\omega)+1}{2(2n+3)\sqrt{(2n+1)(2n+5)} \br{N_+}}\frac{Y}{y_b}, \nonumber
\end{eqnarray}
where $r=0$ or $r=1$ corresponds to absent or present string fluctuations, respectively, $y_b$ denotes the rapidity of the beam, $[-Y,Y]$ is the
range of the acceptance bin, and $s(\omega)$  is the square of the scaled standard deviation of the
fluctuating distribution of strength of the sources (such additional fluctuations are indeed needed to obtain the multiplicity distributions in
the Glauber approach). The results for the $a_{nm}$ coefficients are shown in the right part of Fig.~\ref{fig:anm}. 

With the help of experimental data one may carry out an interesting analysis. The observed number 
of charged hadrons is proportional to the number of sources, $N_{\rm ch}=c N_+$. We may 
then obtain $c$ from matching $a_{11}^{\rm exp}=a_{11}^{\rm mod}$. This yields 
$N_{\rm ch} = 4.7 N_+$. In the CMS acceptance range $\Delta \eta=4.8$, hence we have $dN_{\rm ch}/d\eta \simeq 1 \times N_+$. 
On the other hand, from the multiplicity data $dN_{\rm ch}/d\eta \simeq (3-4) \times N_W$ and $dN_{\rm ch}/d\eta \simeq 1.3 \times Q_W$~\cite{Bozek:2016kpf}, where 
$N_W$ indicates the number of wounded nucleons and $Q_W$ the number of wounded quarks. We note that the number inferred from the matching 
of $a_{11}$ suggests the wounded quark picture, or more generally, a picture with $\sim 3$ constituents (hot-spots) in the nucleon.

\begin{figure}[tb]
\begin{center}
\subfigure{\includegraphics[angle=0,width=.66 \textwidth]{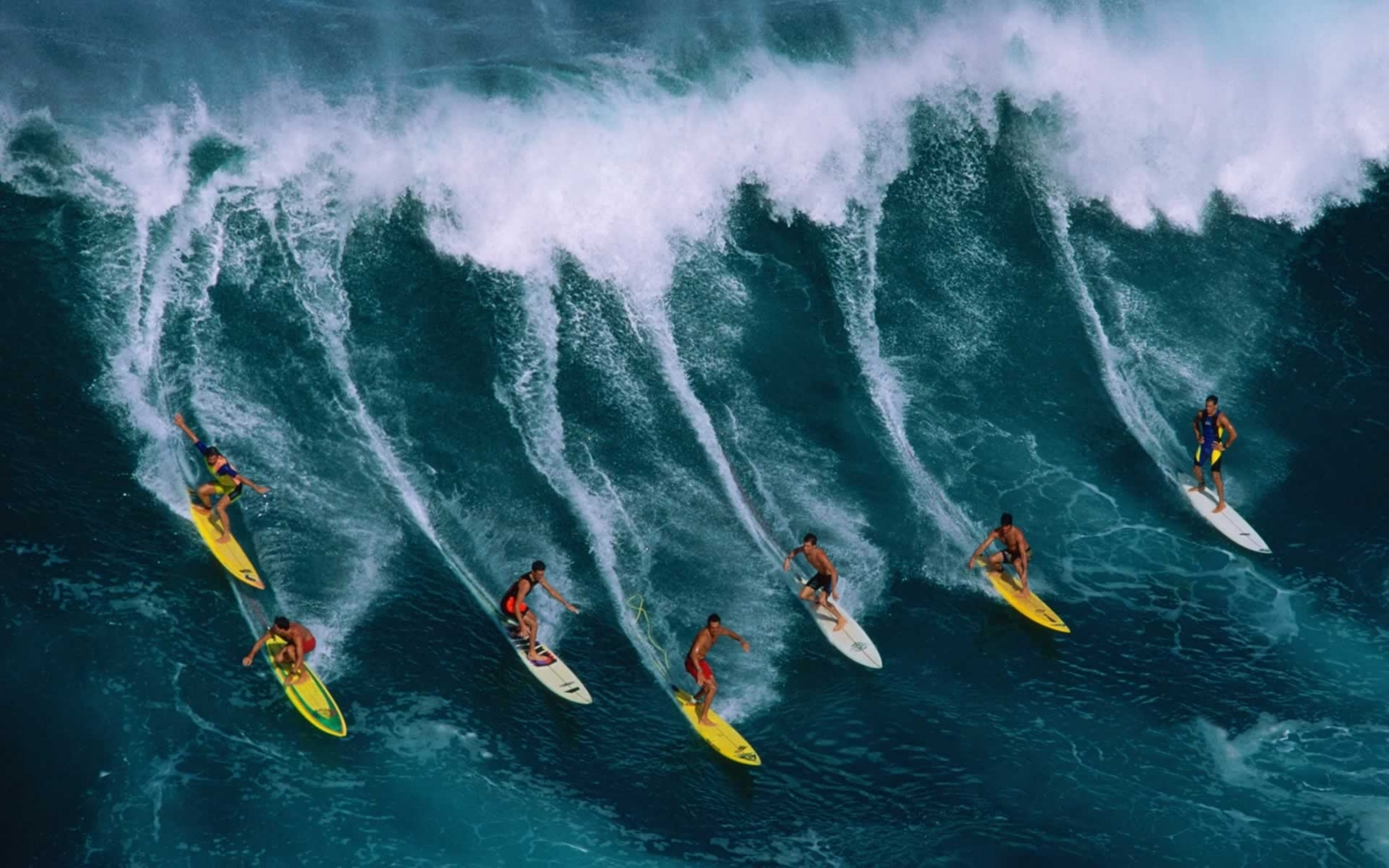}}  \subfigure{\includegraphics[angle=0,width=.31 \textwidth]{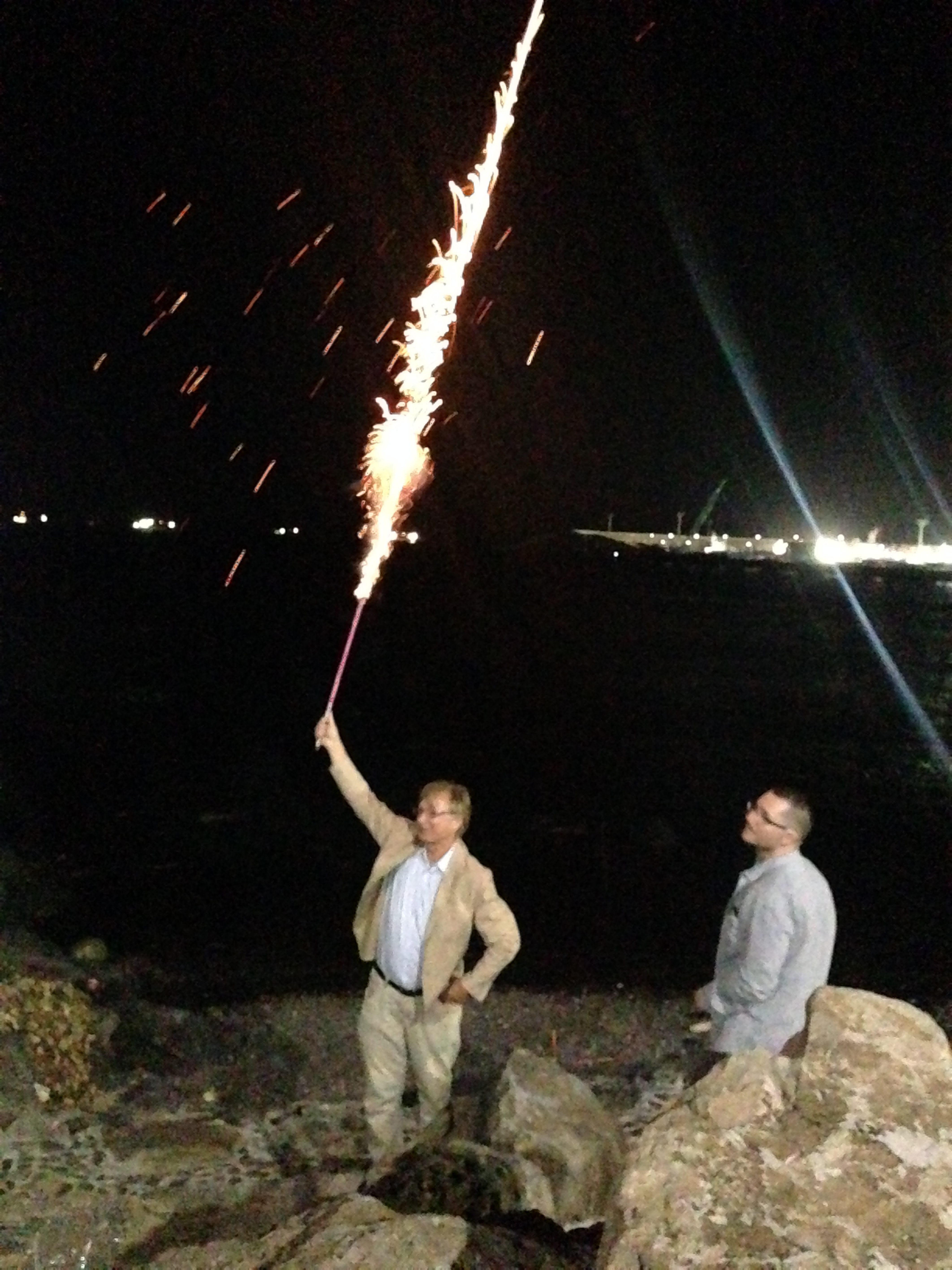}}
\caption{Collectivity of flow (left) and rare non-flow events (right). \label{fig:fun}}
\end{center}
\end{figure}

We conclude this talk by showing two photographs alluding to the collectivity paradigm (left part of Fig.~\ref{fig:fun}), occasionally questioned 
during the meeting. The surfers motion is strongly collimated even if they are 
2~miles away, due to the underlying collective wave. On the other hand, the non-flow events, while present, are quite rare (right part of Fig.~\ref{fig:fun}).

\bigskip \bigskip

Research supported by the Polish National Science Centre grants 2015/17/B/ST2/00101, 2012/06/A/ST2/00390,  and 2015/19/B/ST2/00937.

\eject

\bibliography{hydr}

\end{document}